# Comments on Violations of the Second Law

Elias P. Gyftopoulos
Massachusetts Institute of Technology
77 Massachusetts Avenue
Cambridge, Massachusetts  02139

The authors of Ref. [1] claim to have experimental verification of violations of the second law of thermodynamics based on the assertions: (i) "for large systems and over long times the entropy production rate is necessarily positive"; (ii) Loschmidt's paradox interpreted as indicating that entropy production can be both positive and negative; (iii) and violations of the second law for small systems over short time scales predicted by the fluctuation theorem FT [2].

Neither the claim nor the assertions are correct.  First, without reference to quantum theory, a logically consistent, unambiguous, noncircular, and complete exposition of thermodynamics [3] reveals that any system (both large, and small, including a system of one particle), in any state (both thermodynamic equilibrium, and not thermodynamic equilibrium) has a nonstatistical, instantaneous, property called entropy in the same sense that any state has inertial mass as a property.  Among many theorems about entropy, two are: (a) IF a system experiences an adiabatic reversible process, the entropy is invariant; and (b) IF a system experiences an adiabatic irreversible process, the entropy increases.  The key word here is "IF", and therefore the entropy production rate need not be positive.

Second, if a process obeys the laws of classical mechanics, both the entropy and the entropy production rate are zero.  Accordingly, no proof of either positivity or negativity is required.

Third, in a unified quantum theory of mechanics and thermodynamics it is shown that: (1) the probabilities associated with measurement results must be represented solely by a density operator $\rho$ [4] that corresponds to a homogeneous or unambiguous ensemble (Figure 1); (2) of the myriad of expressions which have been proposed for entropy, only one functional of $\rho$ satisfies all the conditions that must be satisfied by the entropy of thermodynamics [5]; and (3) the entropy is a measure of the quantum-theoretic spatial shape of each particle of a system [6].  So both the FT theorem, and its experimental verification are not and cannot be correct because classical mechanics does not account for thermodynamic phenomena.

# HOMOGENEOUS ENSEMBLE

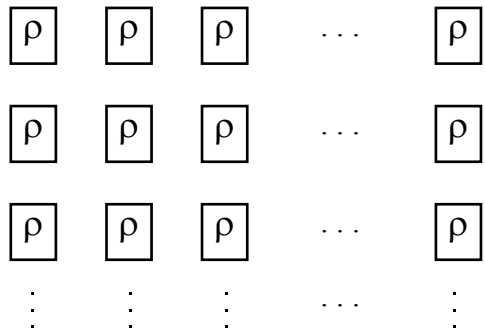

OVERALL DENSITY OPERATOR = $\rho$

Figure 1: Pictorial representation of a homogeneous ensemble. Each of the members of the ensemble is characterized by the same density operator $\rho \geq \rho^2$. It is clear that any conceivable subensemble of a homogeneous ensemble is characterized by the same $\rho$ as the ensemble.